\documentclass[12pt]{article}
\pdfoutput=1
\usepackage{amsmath}
\usepackage{jheppub}
\usepackage{graphicx}
\usepackage{slashed}
\usepackage[thinlines]{easytable}
\usepackage{amssymb}
\usepackage{amsthm}

\usepackage{verbatim,graphicx}
\usepackage{mathtools}
\usepackage{color}
\usepackage[all]{xy}
\usepackage{simplewick}
\usepackage{hyperref}
\usepackage{setspace}
\usepackage{array}
\usepackage{tikz}
\usepackage{tkz-euclide}
\usepackage{enumitem}
\usepackage{bbm}
\usepackage{bbold}

\newcounter{yjcc}

\newcounter{nscc}

\newcommand{\Tr}{{\rm Tr}}

\newcommand{\nn}{\nonumber}

\newcommand{\be}{\begin{eqnarray}}
\newcommand{\ee}{\end{eqnarray}}

\newcommand{\vev}[1]{\left\langle #1\right\rangle}

\newcommand{\bmat}{\left ( \begin{array}{cc} }
	\newcommand{\emat}{\end{array} \right ) }

\newcommand{\eqnum}{\refstepcounter{equation}\textup{\tagform@{\theequation}}}

\usetikzlibrary{shapes.misc}

\tikzset{cross/.style={cross out, draw=black, fill=none, minimum size=2*(#1-\pgflinewidth), inner sep=0pt, outer sep=0pt}, cross/.default={2pt}}

\def\Tr{\textrm{Tr}}

\usetikzlibrary{shapes.misc}

\newcommand*{\rdmathspace}[1][h]{%
\if h#1%
\thinmuskip=2mu\medmuskip=1mu\thickmuskip=3mu\fi%
\if m#1%
\thinmuskip=2.3mu\medmuskip=2mu\thickmuskip=3.5mu\fi%
\if s#1%
 \thinmuskip=3mu minus 0.7mu\medmuskip=4mu minus 2mu\thickmuskip=5mu plus 3.5mu minus 1.5mu\fi%
\if d#1%
\thinmuskip=3mu\medmuskip=4mu\thickmuskip=5mu plus 5mu\fi}

\newcommand{\beq}{\begin{equation}}
\newcommand{\beqs}{\begin{equation*}}
\newcommand{\eeq}{\end{equation}}
\newcommand{\eeqs}{\end{equation*}}

\allowdisplaybreaks

\title{Contact diagrams for chords}
\author{Yiyang Jia}
\affiliation{Department of Particle Physics and Astrophysics, Weizmann Institute of Science, Rehovot, Israel}
\emailAdd{yiyang.jia@weizmann.ac.il}
\date{}
\abstract{
We demonstrate how  contact chord diagrams can arise from certain Fock-space models and compute the corresponding correlation functions using the chord path integral technique.  In particular, our three-point functions are in the right form dictated by conformal symmetry,  and some of our four-point functions match the results of some AdS$_2$ contact Witten diagrams.
}
\begin{document}

\maketitle

\section{Introduction}
The Sachdev-Ye-Kitaev (SYK) model describes a $p$-body all-to-all interaction among $N$ Majorana fermions. The low-energy sector of the SYK model is captured by a Schwarzian action. The Schwarzian dynamics is holographically due to a AdS$_2$ spacetime with a fluctuating cut-off boundary (near-AdS$_2$),  which can be described by a Jackiw-Teitelboim action \cite{kitaev2015,maldacena2016,jensen2016,maldacena2016a,almheiri2014}. The success of the SYK model is largely due to its exact solvablity in the large $N$ limit using $G\Sigma$ bilocal mean field or Schwinger-Dyson equations.    There is another interesting limit  where $p$ scales as $\sqrt{N}$.  This is called the double-scaled limit of the SYK model (DSSYK), and this limit can be solved by chord diagram techniques which are entirely combinatorial \cite{erdos2014, cotler2016, Berkooz:2018qkz,Berkooz:2018jqr,garcia2018c}.  
Recently, an chord path integral was derived in  \cite{Berkooz:2024evs, Berkooz:2024ofm} which allows us to solve  more general models that are not amenable to  $G\Sigma$ approach (also see \cite{Gao:2024lem} for a nice application).  This technique will be crucial for the main development of the current paper.

A good case can be made that one may think of chords as spacetime processes of particles in near-AdS$_2$ geometry, that is, chords that represent matter can be thought of as the spacetime geodesics of the matter particles \cite{Berkooz:2018jqr} (for the thermal field double version see \cite{Lin_2022}).  If this is the case, one would hope to build ``contact'' diagrams of chords to describe contact interactions of the particles, since such interactions are generically present in a tentative parent theory of near-AdS$_2$ gravity,  such as Type IIB strings on AdS$_5\times S^5$.  However, it is not feasible to build such diagrams in the original DSSYK model.   Lately it has been pointed out that the DSSYK model belongs to a much larger class of models characterized by random fluxes in the Fock space, which all have the same leading-order behaviour \cite{Berkooz_2024,berkooz2023parisis, Jia:2024tii}. We will demonstrate how to construct chord $n$-point contact diagrams using these generalized models and provide some prototype examples.  Moreover, using the recently developed chord path integral techniques,  we compute the  correlation functions of the three-point and four-point contact diagrams. For the three-point function,  the result gives what can be expected from conformal symmetry. More interestingly, some of the four point functions we compute exactly match the results of some AdS$_2$ Witten diagrams \cite{Bliard_2022}.

\section{No contact diagrams for double-scaled SYK}
\begin{figure}
    \centering
    \includegraphics[width=0.25\linewidth]{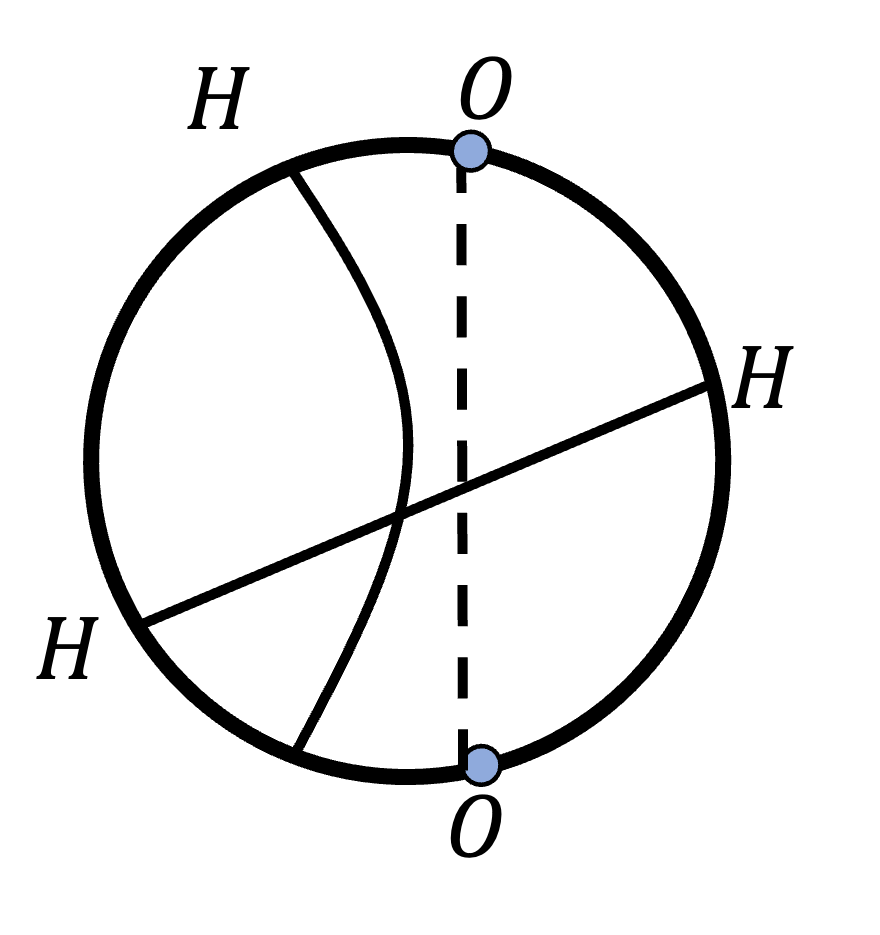}
    \caption{A chord diagram with a two-point insertion. This diagram contributes to $M_{3,1}$ and evaluates to $q^{1+\Delta}$.}
    \label{fig:twopt}
\end{figure}

The $p$-body SYK model is defined by the Hamiltonian 
\begin{equation}
    H = \sum_{I} J_I \Psi_I
\end{equation}
where  $I$ is an index set 
\begin{equation}
    I = \{i_1, i_2, \ldots, i_p\} \quad \text{with} \quad 1\leq i_1<i_2<\cdots<i_p \leq N,
\end{equation}
and $J_I$ are i.i.d Gaussian-distributed random numbers with zero mean and variance $\binom{N}{p}^{-1}$. Furthermore, 
\begin{equation}
    \Psi_I =i^{p/2}\psi_{i_1}\psi_{i_2}\cdots\psi_{i_p} 
\end{equation}
with $\psi_i$ being Majorana fermions. The DSSYK is the model where one takes the limit 
\begin{equation}
    p^2/N = \lambda \text{ fixed}, \quad N \to \infty.
\end{equation}
One can compute all moments in this limit as a polynomial of $q$:
\begin{equation}
    q = \vev{(-1)^{|I_1\cap I_2|}} = \binom{N}{p}^{-2} \sum_{I_1, I_2} (-1)^{|I_1\cap I_2|} = e^{-2\lambda}
\end{equation}
and the $2k$th moments can be expressed as 
\begin{equation} \label{eqn:moments}
   M_{2k} = \sum_\text{chord diagrams with $k$ chords} q^{\text{number of chord intersections}}.
\end{equation}
It turns out one can explicitly sum over the moments to obtain observables such as the spectral density, which is given by the $q$-Gaussian ($q$-Hermite) density.  
The natural definition of probe operators in DSSYK is 
\begin{equation}\label{eqn:DSSYKprobe}
    O = \sum_{I} \tilde J_{\tilde I} \Psi_{\tilde I}
\end{equation}
where  $\tilde J_{\tilde I}$ are i.i.d Gaussian independent of $J_{I}$ and $\tilde I$ is a index set of length $\tilde p$.   The moments of a two-point insertion is given by 
\begin{equation}\label{eqn:twopt-mom}
\begin{split}
        M_{k_1,k_2}:=& \text{(Hilbert space dim)}^{-1}\ \Tr(H^{k_2}O H^{k_1}O)\\ =&  \sum_{\text{CD}_{k_1,k_2}} q^{\# \text{ of $H-H$ intersections}} q^{\Delta (\# \text{ of $O-H$ intersections})}
\end{split}
\end{equation}
where $\Delta = \tilde p/p$ and $\text{CD}_{k_1,k_2}$ denote all chord diagrams with $k_1$ of $H$-chord roots on one side of the $O$-chord and $k_2$ of $H$-chord roots on the other side.  See figure \ref{fig:twopt} for an example. 
It has been shown that in the near-CFT (NCFT) limit 
\begin{equation}\label{eqn:ncftlimit}
    \lambda \to 0^+,\quad  \lambda^{-3/2}>\beta > \lambda^{-1/2}
\end{equation}
we can recover the Schwarzian density,  conformal two-point functions and the four-point functions (especially the out-of-time-ordered one) predicted by the universal near-AdS$_{2}$ dynamics (Jackiw-Teitelboim gravity).  This is the meaning of NAdS$_{2}$/NCFT$_{1}$ duality.  Our current paper will only discuss direct interactions among matters,  so the distinction between near-AdS$_2$ and AdS$_2$ does not really matter and we will not take the effort to distinguish them in language. 

It is not feasible to build contact diagrams using the probes \eqref{eqn:DSSYKprobe}.  On the gravity side, the AdS$_2$ geometry is really part of AdS$_2\times M$ geometry where $M$ describes the shape of the black hole horizon.  Now for contact diagrams, particles must meet at a point on $M$,  which must bring a kinematic suppression factor inversely proportional to (some power of) the size of $M$ \cite{iliesiu2020}, and the size of $M$ is proportional to $N$ by virtue of Bekenstein-Hawking.  For a three point function, the suppression should be $1/\sqrt{N}$, and for a four point function it is $1/N$.  It is important that the contact diagrams built out of our probes reproduce the correct suppression factors.

Let us look at the three point function of $O$,  the very first problem is that three $J_{\tilde I}$ average to zero. Even if we ignore the averaging for a moment,  the only way to have a nonzero contribution is to look for index sets $\tilde I_1, \tilde I_2$ and $\tilde I_3$ such that 
\begin{equation}
    \Tr(\Psi_{\tilde I_1}\Psi_{\tilde I_2}\Psi_{\tilde I_3}) \neq 0.
\end{equation}
The nonzero contribution comes from where we split each of $I_i$ into a pair of index sets with length $\tilde{p}/2$,  and then pair-wise match these six sets.   In the double scaled limit, summing over these configurations will not produce the correct suppression factor.  

Another obstacle, which is more heuristic than quantitative,  is that the diagrams we are seeking from the DSSYK side should not be reducible to pairs of some more elementary objects, or else it is hard to imagine how they can be ``contact''.  One can try variants of DSSYK, but we find it hard to overcome these obstacles. One exception is the double-scaled sparse SYK model \cite{swingle2020,garcia2021}, which contain diagrams that could be interpreted as  four-point contact diagrams \cite{Berkooz_2024},  but it still cannot produce the more basic three-point contact diagrams. 

\section{Contact diagrams from Fock-space models}
It was pointed out earlier that the DSSYK belongs to a much larger class of models which all have the same leading-order behaviour \cite{berkooz2023parisis, Berkooz_2024}. The general Hamiltonian is of the form 
\begin{equation}
    H = \sum_{I}(M_I+ M_I^\dagger) 
\end{equation}
and the form of the index set $I$ does not matter much  as long as it is chosen in such a way that the leading moments of $H$ are given by Wick contractions, e.g. it does not need to be $p$-local, it is completely fine to have $I$ in the form of nearest neighbors (see \cite{Jia:2024tii} for examples).  The important thing is that the $M_I$ operators satisfy a magnetic algebra
\begin{equation}\label{eqn:magAlg}
    M_I M_J = e^{iF_{IJ}} M_J M_I 
\end{equation}
and the fluxes $F$ satisfies 
\begin{enumerate}
    \item $F$ with different pairs of antisymmetric $[IJ]$ are independently and identically distributed.
    \item $\vev{\sin F} = 0$ and $q:=\vev{\cos F}$ is a finite tunable parameter.
\end{enumerate}
The DSSYK itself satisfy these properties with $M_I = M_I^\dagger = J_I\Psi_I, \ F_{IJ} = 0, \pi$ (for more details see \cite{Berkooz_2024}). All such models gives the same moments as the DSSYK moments \eqref{eqn:moments} and \eqref{eqn:twopt-mom} (and higher-point correlators) at leading order.  There is however another construction that directly starts with $N$ Fock-space oscillators $S_i^\pm$ and a center element/charge operator $S_i^3$ ($i=1,2,\ldots, N$), which satisfy 
\begin{equation}
    [S^3, S^\pm] = \pm S^\pm.
\end{equation}
From this we can build the fluxed operators 
\begin{equation}
    T_\mu^\pm = S^\pm_\mu \prod_{\rho \neq \mu} e^{\pm \frac{i}{2} F_{\mu \rho} S_\rho^3}.
\end{equation}
which satisfy 
\begin{equation}\label{eqn:TmagAlg}
    T_\mu^+ T_\nu^+ = e^{iF_{\mu\nu}}T_\nu^+T_\mu^+,  \quad T_\mu^+ T_\nu^- = e^{-iF_{\mu\nu}}T_\nu^-T_\mu^+
\end{equation}
exactly in the form of equation \eqref{eqn:magAlg}.  To keep the setting as general as possible, we allow $F$ to take values on the whole real line, i.e. we allow for noncompact gauge fields.
The form of the Hamiltonian is quite general, and a very simple one can be 
\begin{equation}\label{eqn:basicHami}
    H = \frac{\text{const}}{\sqrt{N}} \sum_{\mu=1}^N D_\mu :=\frac{\text{const}}{\sqrt{N}} \sum_{\mu=1}^N (T_\mu^+ + T_\mu^-) 
\end{equation}
where the constant is chosen such that $\Tr H^2 = $ Hilbert space dimension.  If we plot the Fock-space graph of the Hamiltonian \eqref{eqn:basicHami}, we get a $N$-dimensional hypercubic lattice.  And if the dimension of the representation of $S^\pm$ and $S^3$ is $L$,  then the hypercubic lattice has $L$ sites along each direction. Taking trace picks up all the lattice paths that return to the starting point (loops).
The leading moments are given by configurations of distinct pairs, for example 
\begin{equation}
    \Tr H^4 = \frac{\text{const}^2}{N^2} \sum_{\mu\neq \nu }\Tr(D_\mu D_\mu D_\nu D_\nu+ D_\mu D_\nu D_\mu D_\nu+ D_\mu D_\nu D_\nu D_\mu).
\end{equation}
After averaging, at leading order all moments are given by equation \eqref{eqn:moments}, namely the DSSYK moments.   

What about subleading order?  This happens when we have an index being repeated more than twice,  for example, we can have 
\begin{equation}\label{eqn:hami-sub}
    \Tr H^3 = \frac{\text{const}^3}{N^{3/2}}\sum_{\mu} \Tr(D_\mu D_\mu D_\mu),
\end{equation}
which may or may not be zero depending on the form of $T_\mu^\pm$ we choose (see next section).  This form of moments exactly overcomes the previously mentioned obstacles:  it has the right suppression factor $1/\sqrt{N}$ and it is not reducible to pairs of more elementary objects: this is a candidate ``contact'' contribution.  However, we will only concern ourselves with contact diagrams of probe operators instead of the Hamiltonian.\footnote{The moment \eqref{eqn:hami-sub} only contributes to the partition function but not the three-point function.} In the same spirit as DSSYK probes \eqref{eqn:DSSYKprobe},  we define the probe operators as 
\begin{equation}
    O = \frac{\text{const}}{\sqrt{N}} \sum_{\mu=1}^N \tilde D_\mu :=\frac{\text{const}}{\sqrt{N}} \sum_{\mu=1}^N (\tilde T_\mu^+ + \tilde T_\mu^-) 
\end{equation}
where $\tilde T_\mu^\pm$ is defined in the same way as $T_\mu$ but with another set of fluxes $\tilde F_{\mu\nu}$, which we again require to be i.i.d distributed for distinct antisymmetric pairs $[\mu\nu]$,  but we do not require $\tilde F_{\mu\nu}$ to be independent of $F_{\mu\nu}$.  For example,  $\tilde F_{12}$ needs to be independent of $\tilde F_{23}$, but $\tilde F_{12}$ may well be correlated with $F_{12}$.  As we shall see, such correlations are important if we want to match the four-point chord contact diagrams with some of the four-point Witten diagrams.  We shall remark that such requirements are in place mainly for the chord diagram techniques to work.  The true form of the distributions may well be more general than our requirement as far as the physics is concerned.  With the extra set of fluxes, we additionally have  

    \begin{equation}
    T_\mu^\pm \tilde T_\nu^+ = e^{\pm i \frac{F_{\mu\nu}+\tilde F_{\mu\nu} }{2}}T_\nu^+T_\mu^+,  \quad T_\mu^{\pm} \tilde T_\nu^- =  e^{\mp i \frac{F_{\mu\nu}+\tilde F_{\mu\nu} }{2}}\tilde T_\nu^-T_\mu^+, 
\end{equation}
and we define 
\begin{equation}\label{eqn:q_and_delta}
    q^\Delta := \vev{\cos \frac{F+ \tilde F}{2}} \implies \Delta = \frac{\log \vev{\cos \frac{F+ \tilde F}{2}}}{\log \vev{\cos F}}
\end{equation}
We can achieve the conformal limit \eqref{eqn:ncftlimit} by making the distribution of $F, \tilde F$ more and more concentrated on $F =0, \tilde F =0$, and in this case 
\begin{equation}
    \Delta = \frac{\vev{(F+\tilde F)^2}}{4\vev{F^2}}
\end{equation}
is the conformal dimension of our probe. 

In summary, for a Euclidean $n$-point function 
\begin{equation}
    \vev{O_{n}(\tau_n)O_{n-1}(\tau_{n-1}) \cdots O_1(\tau_1) },
\end{equation}
our candidate $n$-point contact contributions are of the form
\begin{equation}\label{eqn:n-point-contact-mom}
   \sim \frac{1}{N^{(\sum_{i}k_i + n)/2}} \sum_\mu\Tr(H^{k_n}\tilde D_\mu^{(n)} H^{k_{n-1}}\tilde D_\mu^{(n-1)} \cdots H^{k_1}\tilde D_\mu^{(1)})
\end{equation}
where each $\tilde D_\mu^{(i)}$ in principle can be defined by a distinct flux $\tilde F^{(i)}$.
\subsection{Three-point  diagrams}\label{sec:threept} 
First we seek a construction that has nonzero  $\langle \Tr \tilde D_\mu  \tilde D_\mu \tilde D_\mu \rangle$. As mentioned earlier, we can visualize the trace as picking up loops on the hypercubic lattice.  How can we form a three-step loop? The natural choice is to take a lattice with three sites along each direction and compactify all the directions. 
In term of Fock space oscillators, we choose the representation 
\begin{equation}
    S_3 = \begin{pmatrix}
    1&0&0\\
    0&0&0\\
    0&0&-1
    \end{pmatrix},\  S^+ =  \begin{pmatrix}
    0&1&0\\
    0&0&1\\
    0&0&0
    \end{pmatrix},\  S^- =  \begin{pmatrix}
    0&0&0\\
    1&0&0\\
    0&1&0
    \end{pmatrix}
\end{equation}
which reflect the fact that we are using a size-three lattice.  To implement compactification, consider 
\begin{equation}
    P^+ =   \begin{pmatrix}
    0&1&0\\
    0&0&1\\
    1&0&0
    \end{pmatrix} = S^+ +(S^-)^2, \quad P^- = (    P^+)^\dagger.
\end{equation}
and implement the magnetic operators as 
\begin{equation}
        T_\mu^\pm = P^\pm_\mu \prod_{\rho \neq \mu} e^{\pm \frac{i}{2} F_{\mu \rho} S_\rho^3}.
\end{equation}
However, to retain the magnetic algebra in equation \eqref{eqn:TmagAlg},  we will need 
\begin{equation}
  e^{i \frac{F}{2} S^3} P^\pm  e^{-i \frac{F}{2} S^3}  =   e^{\pm i \frac{F}{2} } P^\pm.
\end{equation}
which only works if the fluxes are rational:\footnote{This has a well-known analog in the lattice Landau/Hofstader problem, where periodic boundary condition on the gauge potentials enforces rational fluxes.}
\begin{equation}
    F = \frac{4n\pi}{3}, \quad n\in \mathbb{Z}
\end{equation}
and the same goes for $\tilde F$.  
\begin{figure}
    \centering
    \includegraphics[scale=0.7]{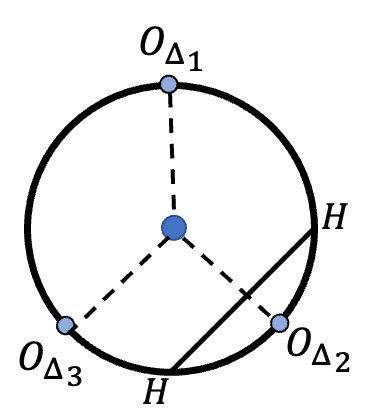}
    \caption{Chord diagram for a contact three-point function, which has an overall scaling of $1/\sqrt{N}$. The dashed lines represent the probes, which meet at the middle of the disk and form a three-leg structure. Each time an $H$-chord crosses a leg representing $O_{\Delta_i}$, we get a factor of $ q^{\Delta_i}$. The diagram we plotted evaluates to $ q^{\Delta_2} /\sqrt{N}$.}
    \label{fig:threePtChord}
\end{figure}
To be explicit with the normalization, we choose the Hamiltonian to be 
\begin{equation}
    H = \frac{1}{\sqrt{6N}}\sum_\mu ( T_\mu^+ +  T_\mu^-),
\end{equation}
so that $3^{-N} \langle \Tr H^2\rangle=1$. The same goes for probes $O_i$ only with $F$ replaced by $\tilde F^{(i)}$. Note that 
\begin{equation}
   3^{-N} \vev{\Tr (\tilde T_\mu^{(1)+} \tilde T_\mu^{(2)+} \tilde T_\mu^{(3)+}) }\propto 
   \left(2 \vev{ \cos \frac{\tilde F^{(1)} + \tilde F^{(2)}+ \tilde F^{(3)} }{2}} +1\right)^{N-1} /3^N.
\end{equation}
To avoid an undesirable exponential suppression, we require 
\begin{equation}
    \sum_{i=1}^3 \tilde F^{(i)} \in 4\pi \mathbb{Z}.
\end{equation}
One can derive the general chord rules for a three-point contact contribution
\begin{align}
    &3^{-N}   \vev{ \Tr H^{k_3} O_3  H^{k_2} O_2 H^{k_3} O_1}_\text{contact} \nn \\
    =& \frac{1}{\sqrt{N}}  \sum_\text{CD} q^{\text{\# $H$-$H$ int.}} q^{\Delta_1(\text{\# $H$-$O_1$ int.)} +\Delta_2(\text{\# $H$-$O_2$ int.)}+\Delta_3(\text{\# $H$-$O_3$ int.)}},\label{eqn:3ptChordRules}
\end{align}
where again  
\begin{equation}
    q = \vev{\cos F}, \quad q^{\Delta_i} = \vev{\cos\frac{F + \tilde F^{(i)}}{2}},
\end{equation}
and $\sum_\text{CD}$ means summing over all possible chord diagrams with a three-point contact insertion.   A diagramatic representation of a moment  is given by figure  \ref{fig:threePtChord} and the meaning of ``\# $H$-$O_i$ int.'' should be evident from the diagram.

\subsection{Four-point  diagrams}\label{sec:fourpt}
\begin{figure}
    \centering
    \includegraphics[width=0.75\linewidth]{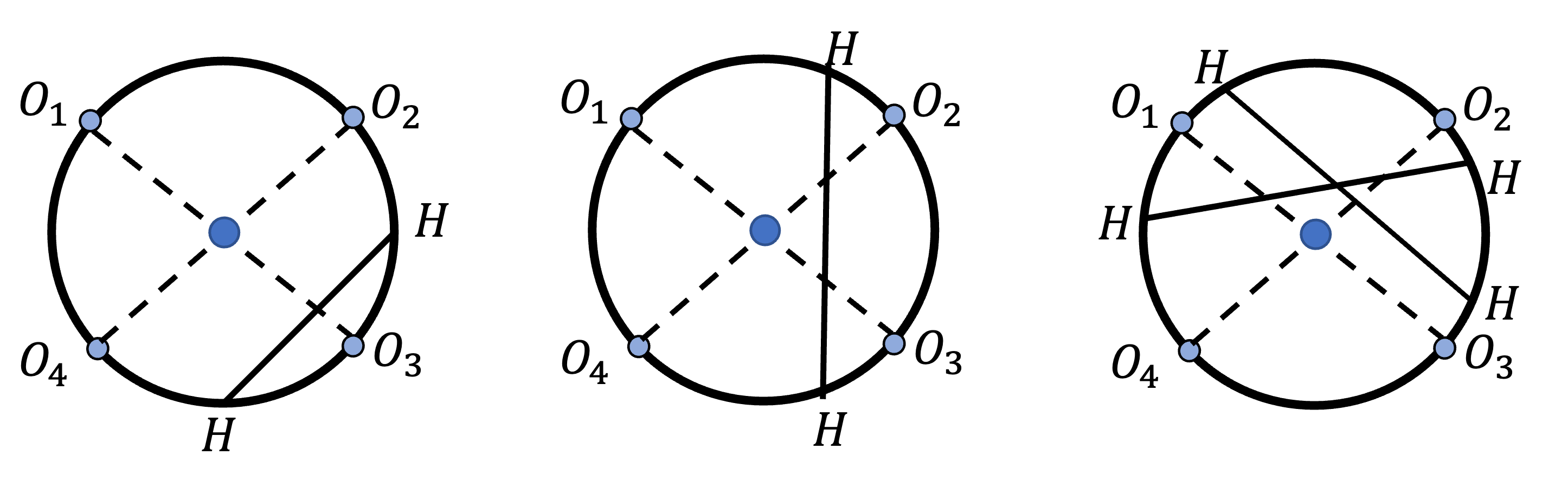}
    \caption{Chord diagrams with a contact four-point insertion.  Left diagram equals $q^{\Delta_3}/N$. Middle diagram equals $q^{\Delta_{23}}/N$. Right diagram equals $q^{1+\Delta_2 + \Delta_{12}}/N$.}
    \label{fig:4ptchord}
\end{figure}

As first observed in \cite{Berkooz_2024},  four-point contact diagrams already appear on the simplest size-two lattice with no compactification, where 
\begin{equation}
 S_3 = \frac{1}{2}\begin{pmatrix}
        1&0 \\ 0&-1
    \end{pmatrix},\quad  S^+ = \begin{pmatrix}
        0&1 \\ 0&0 
    \end{pmatrix}, \quad S^- = \begin{pmatrix}
        0&0 \\ 1&0 
    \end{pmatrix},
\end{equation}
Let us deal with this simplest scenario and later comment on larger lattice sizes.  The normalization of the Hamiltonian is\footnote{This Hamiltonian turns out to be the same as a model first proposed by Parisi in a quite different context \cite{Parisi:1994jg,marinari1995} and its connection to the DSSYK was first noticed by \cite{Jia_2020}.} 
\begin{equation}
    H = \frac{1}{\sqrt{N}}\sum_{\mu=1}^N D_\mu = \frac{1}{\sqrt{N}}\sum_{\mu=1}^N (T_\mu^+ + T_\mu^-),
\end{equation}
and similarly for the probes.  
For the four-point function 
\begin{equation}
    \vev{O_4(\tau_4)O_3(\tau_3)O_2(\tau_2) O_1(\tau_1)},
\end{equation}
the contact contribution to the moments is given by 
\begin{equation}\label{eqn:4point-contact-mom}
    \frac{1}{N^{2+\sum_{i}k_i/2}} \sum_\mu\Tr(H^{k_4}\tilde D_\mu^{(4)} H^{k_{3}}\tilde D_\mu^{(3)}  H^{k_2}\tilde D_\mu^{(2)} H^{k_1}\tilde D_\mu^{(1)})
\end{equation}
Since in this case $(\tilde T_\mu^+)^2 = (\tilde T_\mu^-)^2 = 0$,  this contribution can be broken up to two pieces:
\begin{equation}
    \Tr(\cdots \tilde T_\mu^+ \cdots \tilde T_\mu^- \cdots \tilde T_\mu^+ \cdots\tilde T_\mu^- ) + \Tr(\cdots \tilde T_\mu^-\cdots  \tilde T_\mu^+ \cdots \tilde T_\mu^- \cdots \tilde T_\mu^+  ) 
\end{equation}
Taking trace would give a 
\begin{equation}
    \vev{\cos \frac{\tilde F^{(1)} - \tilde F^{(2)}  + \tilde F^{(3)} - \tilde F^{(4)}}{2}}^{N-1}
\end{equation}
exponential factor in the contribution, therefore we require the constraint
\begin{equation}\label{eqn:constraint2latt}
    \tilde F^{(1)} - \tilde F^{(2)}  + \tilde F^{(3)} - \tilde F^{(4)} \in 4\pi \mathbb{Z}. 
\end{equation} 
With this constraint imposed, we get the chord rules:
\begin{equation}\label{eqn:4ptcontact-mom-result}
    \frac{1}{N}\sum q^{\# H-H \text{ int.}} q^{\Delta_i \text{( $\# H-O_i$ int.)}} q^{\Delta_{12} \text{( $\# H-O_1-O_2 $ int.)}}q^{\Delta_{23} \text{( $\# H-O_2-O_3 $ int.)}}
\end{equation}
where 
\begin{equation}\label{eqn:qfactors-4pt}
    q^{\Delta_i} = \vev{\cos \frac{F + F^{(i)}}{2}},  \quad q^{\Delta_{ij}} = \vev{\cos \frac{ F^{(i)}- F^{(j)}}{2}},
\end{equation}
and $\# H-O_1-O_2 $ int. means the number of $H$-chords that cross both $O_1$ and $O_2$ simultaneously. See figure \ref{fig:4ptchord} for examples.
Note due to the constraint \eqref{eqn:constraint2latt} we have $\Delta_{12}=\Delta_{34}$ and $\Delta_{23} = \Delta_{14}$. 

\subsection{More general situations}\label{sec:general-lattice}
Equation \eqref{eqn:4ptcontact-mom-result} is our simplest example of a contact four-point moments.  By now it should be clear how this works in more general situations. For example,  the size-three periodic lattice defined in section \ref{sec:threept} also contains four-point contact contributions, however in this case we also need to consider contributions from $++--$, $--++$, $+--+$ and $-++-$ configurations (here $\pm$ refers to the superscript on the $\tilde T_\mu^\pm$ operators).  If we use a size-four periodic lattice (where fluxes can only take values in $\pi \mathbb{Z}$), we can also have contributions from $++++$ and $----$. Moreover, it is clear the size-two lattice already contains contact contributions of arbitrary $2m$-point, and size-three periodic lattice contains contact contributions of arbitrary $n$-point.  For example, a five-point contact contribution can be realized on the size-three periodic lattice as $++++-$.  However one may still wish to utilize larger lattices to take advantage of more flexibilities.

\section{Computing the correlation functions}
\subsection{Chord path integral}\label{sec:chord-path-int}
Let us give a quick recap on the chord path integral technique developed in \cite{Berkooz:2024evs, Berkooz:2024ofm}.  In the $q = e^{-2\lambda } \to 1^-$ limit, all observables that can be built from chords are controlled by a partition function 
\begin{equation}
    Z = \int \mathcal{D}n  \exp\left(-\frac{1}{\lambda} S[n]\right)
\end{equation}
where the action $S$ is defined as
\begin{equation}
\begin{split}
        S[n] = \frac{1}{4} \int_{0}^{\beta} d\tau_a \int_{0}^{\beta} d\tau_b\int_{\tau_a}^{\tau_b} d\tau_c\int_{\tau_b}^{\tau_a} d\tau_d \ n(\tau_a, \tau_b)n(\tau_c, \tau_d)\\
        +  \frac{1}{2} \int_{0}^{\beta} d\tau_a \int_{0}^{\beta} d\tau_b\ n(\tau_a, \tau_b)[\log (n(\tau_a, \tau_b)) -1]
\end{split}
\end{equation}
where $n(\tau_a, \tau_b)$ is the chord density, i.e.,  $n(\tau_a, \tau_b)d\tau_a d\tau_b$ is the number of chords connecting the intervals $[\tau_a, \tau_a+d\tau_a]$ and $[\tau_b, \tau_a+d\tau_b]$. If $\tau_a <\tau_b$, the integral $\int_{\tau_b}^{\tau_a}$ is interpreted as 
\begin{equation}
  \int_{\tau_b}^{\tau_a}:=  \int_{\tau_b}^{\tau_a + \beta}, \quad \tau_a <\tau_b
\end{equation}
with all the corresponding integrands periodically extended.  The action has a saddle point most conveniently expressed in terms of a new function 
\begin{equation}\label{eqn:ng-relation}
    g(\tau_a,\tau_b):= - \int_{\tau_a}^{\tau_b} d\tau \int_{\tau_b}^{\tau_a} d\tau' \ n(\tau, \tau') \quad \implies  \quad n(\tau_a, \tau_b) = -\frac{1}{2}\partial_{\tau_a}\partial_{\tau_a}g(\tau_a,\tau_b),
\end{equation}
and at the saddle  
\begin{equation}\label{eqn:g-saddle}
    \exp(g(\tau_a, \tau_b)) = \frac{\cos^2 \left(\frac{\pi v}2\right)}{\cos^2 \left[\frac{\pi v}2\left(1-\frac{2|\tau_b-\tau_a|}{\beta
    }\right)\right]},  \quad \beta = \frac{\pi v}{\cos \frac{\pi v}{2}}.
\end{equation}
In the low-temperature limit,  this just gives 
\begin{equation}
    \exp(g(\tau_a, \tau_b))  = \frac{1}{\tau_{ab}^2}, \quad \tau_{ab}:= \tau_b -\tau_a.
\end{equation}
The moments of a two-point insertion is given by equation \eqref{eqn:twopt-mom}. The resulting time-ordered two-point function is given by the exponentiation of the two-point moments.  In the prescription of chord path integral,  we compute the exponential of number of $H$-chords  that intersect with $O$-chord which is located on the thermal circle at $\tau_1$ and $\tau_2$:
\begin{equation}
    \vev{O(\tau_2)O(\tau_1)} = \vev{e^{-\Delta \int_{\tau_1}^{\tau_2} d\tau \int_{\tau_2}^{\tau_1} d\tau' \ n(\tau, \tau') }} = \vev{e^{\Delta g(\tau_1,\tau_2)}},
\end{equation}
where the $\langle\cdots\rangle$ on the right denotes the path-integral averaging using the action $S[n]$. 
Therefore, at the saddle point the two-point function is simply equation \eqref{eqn:g-saddle} raised to the power of $\Delta$, and at the zero-temperature limit it is simply 
\begin{equation}
     \vev{O(\tau_2)O(\tau_1)}  = \frac{1}{|\tau_{12}|^{2\Delta}}.
\end{equation}
\subsection{Contact three-point functions}\label{sec:threeptfunc}
We wish to evaluate $\vev{O_3(\tau_3)O_2(\tau_2)O_1(\tau_1)}_\text{contact}$. Since path integral evaluates the time-ordered correlation, for simplicity let us assume 
\begin{equation}
    \tau_3>\tau_2>\tau_1.
\end{equation}
Following the chord path integral prescription,  the three-point function that arises from the contact contribution \eqref{eqn:3ptChordRules} is 
\begin{align}
       &\vev{O_3(\tau_3) O_2(\tau_2)O_1(\tau_1)}_\text{contact} \nn \\
   =& \frac{1}{\sqrt{N}}\left\langle\exp\left[-\Delta_1 \int_{\tau_1}^{\tau_2} d \tau\int_{\tau_3}^{\tau_1}d \tau' \ n(\tau,\tau') \right. \right. \\  & \qquad \quad \left. \left. 
 - \Delta_2 \int_{\tau_2}^{\tau_3}d \tau \int_{\tau_1}^{\tau_2}d \tau'  n(\tau,\tau')- \Delta_3 \int_{\tau_3}^{\tau_1}d \tau \int_{\tau_3}^{\tau_2}d \tau'  n(\tau,\tau')\right] \right\rangle \nn
\end{align}
where $\Delta_i$ follows the same form as that of equation \eqref{eqn:q_and_delta}.
Using the relation \eqref{eqn:ng-relation}, we have 
\begin{equation} \label{eqn:ngint1}
    \int_{\tau_1}^{\tau_2} d \tau \int_{\tau_3}^{\tau_1}d \tau'n(\tau,\tau') = \frac{1}{2}\left[ g_{23} - g_{12}-g_{13}\right].
\end{equation}
where $g_{ij}:= g(\tau_i,\tau_j)$.  This gives 
\begin{align}
        & \vev{O_3(\tau_3) O_2(\tau_2)O_1(\tau_1)}_\text{contact} \nn \\
         =& \frac {1}{\sqrt{N}} \vev{\exp\left[\frac{\Delta_1 +\Delta_2-\Delta_3}{2}g_{12}+ \frac{\Delta_1 +\Delta_3-\Delta_2}{2} g_{13}+\frac{\Delta_2 +\Delta_3-\Delta_1}{2} g_{23}\right]} \nn \\
          =&
   \frac{1}{\sqrt{N}} \frac{1}{|\tau_{12}|^{\Delta_1+\Delta_2-\Delta_3} |\tau_{13}| ^{\Delta_1+\Delta_3-\Delta_2} |\tau_{23}|^{\Delta_2+\Delta_3-\Delta_1}},
\end{align}
where in going to the third line we have taken the saddle point and then the zero-temperature limit. This is exactly what we expect from conformal symmetry. For the set of operators we provided, the OPE coefficients are one,  it would be interesting to construct examples with more general OPE coefficients which we will not do in this work.

\subsection{Contact four-point functions  and comparison with Witten diagrams}\label{sec:fourptfunc}
Let us evaluate the simplest four-point contact correlation $\vev{O_4(\tau_4)O_3(\tau_3) O_2(\tau_2) O_1(\tau_1)}_\text{contact} $ that arises from a size-two lattice, namely the correlation functions that arises from the moments \eqref{eqn:4ptcontact-mom-result}.   To simplify time ordering let us work with 
\begin{equation}
    \tau_4 >\tau_3>\tau_2>\tau_1.
\end{equation}
Following the path integral prescription we get 
\begin{equation}
    \vev{O_4(\tau_4)O_3(\tau_3) O_2(\tau_2) O_1(\tau_1)}_\text{contact}= \frac{1}{N} \vev{\exp(- K[n])}  
\end{equation}
where 
   \begin{align}
    K[n]  :=& \Delta_1\int_{\tau_4}^ {\tau_1}\int _{\tau_1}^ {\tau_2} d\tau d\tau' n(\tau, \tau') \nonumber + \Delta_2\int_{\tau_1}^ {\tau_2}\int _{\tau_2}^ {\tau_3} d\tau d\tau' n(\tau, \tau') \\
      &+\Delta_3\int_{\tau_2}^ {\tau_3}\int _{\tau_3}^ {\tau_4} d\tau d\tau' n(\tau, \tau')+ \Delta_4\int_{\tau_3}^ {\tau_4}\int _{\tau_4}^ {\tau_1} d\tau d\tau' n(\tau, \tau')   \\
      &+   \Delta_{23} \int_{\tau_3}^ {\tau_4}\int _{\tau_1}^ {\tau_2} d\tau d\tau' n(\tau, \tau')  + \Delta_{12} \int_{\tau_2}^ {\tau_3}\int _{\tau_4}^ {\tau_1} d\tau d\tau' n(\tau, \tau'),\nonumber 
\end{align}
and by taking the $q\to 1$ limit of equation \eqref{eqn:qfactors-4pt} we have 
\begin{equation}
    \Delta_i = \frac{\vev{(F + \tilde F^{(i)})^2}}{4\vev{F^2}},\quad \Delta_{ij}= \frac{\vev{(\tilde F^{(i)} - \tilde F^{(j)})^2}}{4\vev{F^2}}. 
\end{equation}
Note $\Delta_i$ and $\Delta_{ij}$ are not completely independent quantities, we will get back to this at the end of this section.
In terms of $g$ functions we get 
\begin{align}\label{eqn:4diffFluxesg}
     K[n] =& \frac{\Delta_{12}-\Delta_1-\Delta_2}{2}g_{12} +\frac{\Delta_2+\Delta_4-\Delta_{12}-\Delta_{23}}{2}g_{13} \nonumber \\
     &+  \frac{\Delta_{23}-\Delta_1-\Delta_4}{2}g_{14} +  \frac{\Delta_{23}-\Delta_2-\Delta_3}{2}g_{23} \\
     &+ \frac{\Delta_1+\Delta_3  -\Delta_{12}- \Delta_{23}}{2}g_{24} + \frac{\Delta_{12}-\Delta_3-\Delta_4}{2}g_{34} \nonumber
\end{align}
where we have used the equality
\begin{equation}\label{eqn:ngint2}
     \int_{\tau_2}^ {\tau_3}\int _{\tau_4}^ {\tau_1} d\tau d\tau' n(\tau, \tau') = \frac{1}{2}(g_{12} +g_{34} -g_{13} - g_{24}).
\end{equation}
At the saddle point, the zero-temperature result becomes 
\begin{equation}
    \vev{O_4(\tau_4)O_3(\tau_3) O_2(\tau_2) O_1(\tau_1)}_\text{contact}= \frac{1}{N} {|\tau_{13}\tau_{24}|^{(\sum_i^4\Delta_i) -\Delta_
    {12}-\Delta_
    {13}-\Delta_
    {23}} }\prod_{1\leq i<j\leq 4} {|\tau_{ij}|^{\Delta_{ij} -\Delta_i - \Delta_j }},
\end{equation}
where the  spurious dependence on $\Delta_{13}$  cancels (it must since it never appeared in equation \eqref{eqn:4diffFluxesg}, we are only using it to make the expression look more symmetric). To better see its dependence on the cross ratio 
\begin{equation}
    x := \frac{\tau_{12}\tau_{34}}{\tau_{13}\tau_{24}}, \quad 1-x = \frac{\tau_{14}\tau_{23}}{\tau_{13}\tau_{24}},
\end{equation}
we can go to the limit where $\tau_1 = 0 , \tau_2 =x, \tau_3 = 1, \tau_4= \infty$, which gives ($0<x<1$) 
\begin{equation}
   I(x):= \lim_{\tau_4 \to \infty} |\tau_4|^{2\Delta_4}\vev{O_4(\tau_4)O_3(1) O_2(x)O_1(0)}_\text{contact} = \frac{1}{N}x^{\Delta_{12} -\Delta_1- \Delta_2} (1-x)^{\Delta_{23} - \Delta_{2}-\Delta_3}.
\end{equation}

One may compare our result with the results from AdS$_2$ scalar Witten diagrams \cite{Bliard_2022}. Since we have considerable amount of freedom to choose the values of $\Delta_{ij}$,  we can find cases where the two match. For example we have an exact match for the case $\Delta_1 = \Delta_2 = \Delta_3 =1, \Delta_4 = 2$ if we take the liberty to choose $\Delta_{12} = \Delta_{23}= 2$,  the result is (most easily read off from equation \eqref{eqn:4diffFluxesg}) 
\begin{equation}
      \vev{O_{\Delta= 2}(\tau_4) O_{\Delta= 1} (\tau_3)O_{\Delta= 1}(\tau_2) O_{\Delta= 1}(\tau_1)}_\text{chord contact} =  \frac{1}{N} \frac{1}{\tau_{13}\tau_{14}\tau_{24}^2 \tau_{34}}
\end{equation}
whereas the result from AdS$_2$ scalar Witten diagram is \cite{Bliard_2022} 
\begin{equation}
      \vev{O_{\Delta= 2}(\tau_4) O_{\Delta= 1} (\tau_3)O_{\Delta= 1}(\tau_2) O_{\Delta= 1}(\tau_1)}_\text{Witten contact} = \text{constant} \times \frac{1}{\tau_{13}\tau_{14}\tau_{24}^2 \tau_{34}}.
\end{equation}
And we get a near match for the case $\Delta_1 = \Delta_2 = \Delta_3 = \Delta_4 =1$ if we choose $\Delta_{12} = \Delta_{23}= 1$:
\begin{equation}
      \vev{O_{\Delta= 1}(\tau_4) O_{\Delta= 1} (\tau_3)O_{\Delta= 1}(\tau_2) O_{\Delta= 1}(\tau_1)}_\text{chord contact} =  \frac{1}{N} \frac{1}{\tau_{12}\tau_{23}\tau_{34} \tau_{14}} = \frac{1}{N} \frac{1}{(\tau_{13}\tau_{24})^2} \left[\frac{1}{x} + \frac{1}{1-x}\right]
\end{equation}
whereas 
\begin{equation}
      \vev{O_{\Delta= 1}(\tau_4) O_{\Delta= 1} (\tau_3)O_{\Delta= 1}(\tau_2) O_{\Delta= 1}(\tau_1)}_\text{Witten contact} =  \frac{\text{constant} }{(\tau_{13}\tau_{24})^2} \left[\frac{\log(1-x)}{x} + \frac{\log x}{1-x}\right].
\end{equation}
Some of the results in \cite{Bliard_2022} are reported in terms of the function $I(x)$,  and we can also get some  exact matches  in terms of  $I(x)$.   For $\Delta_{1}=\Delta_{2}=\Delta_{3}=2, \Delta_4 =1 $, if we choose $\Delta_{12} = \Delta_{23}=3$ we get  
\begin{equation}
    I_\text{chord contact}^{2221 } =  \frac{1}{N}\frac{1}{x(1-x)}.
\end{equation}
and 
\begin{equation}
    I_\text{Witten contact}^{2221 } =  \frac{\text{const}}{x(1-x)}
\end{equation}
For $\Delta_{1}=3, \Delta_{2}=1, \Delta_{3}=2, \Delta_4 =1 $, if we choose $\Delta_{12} = \Delta_{23}=3$ we get 
\begin{equation}
    I_\text{chord contact}^{3121 } =  \frac{1}{N}\frac{1}{x}.
\end{equation}
and 
\begin{equation}
    I_\text{Witten contact}^{3121} =  \frac{\text{const}}{x}.
\end{equation}
Other examples such as $I^{2211}, I^{2121}$ do not match.  Another potential issue is that if we use  four identically defined probes at the microscopic level, namely $\tilde F^{(1)} =\tilde F^{(2)} = \tilde F^{(3)} = \tilde F^{(4)}$ (instead of merely having $\Delta_1 =\Delta_2=\Delta_3=\Delta_4$),  we  will have $\Delta_{12}= \Delta_{23} =0$ and lose the ability to match the Witten diagram results. However, we are not too discouraged by these since the construction we have used is just one example out of a large class, and it is remarkable enough that several Witten diagrams have exact matches even in the simplest model we provide.

One may also worry if the microscopic definitions of $\Delta_{i}$ and $\Delta_{ij}$ allow  us to choose the values we chose.  Let us denote 
\begin{equation}
 a_i  = \frac{F + \tilde F^{(i)}}{2\sqrt{\vev{F^2}}}, \qquad i =1,2,3
\end{equation}
then in the NCFT limit we have 
\begin{align}
    &\Delta_i = \vev{a_i^2},\qquad i = 1,2,3, \\
    & \Delta_4 = \vev{(a_1- a_2 +a_3)^2}, \\
    & \Delta_{12} = \vev{(a_1- a_2 )^2},\\
    & \Delta_{23} = \vev{(a_2- a_3 )^2}.
\end{align}
Therefore 
\begin{equation}
\begin{split}
     &\Delta_{12} =\Delta_1+\Delta_2 -2 \vev{a_1a_2}, \quad \Delta_{23} =\Delta_2+\Delta_3 - 2\vev{a_2a_3} \\
     &\Delta_4 = \Delta_{12} + \Delta_{23}- \Delta_2 + 2 \vev{a_1a_3}.
\end{split}
\end{equation}
The only constraint  is the semi-positivity of the three-by-three covariance matrix $\vev{a_ia_j}$. That is, the matrix 
\begin{equation}
    \begin{pmatrix}
        \Delta_1 & \frac{\Delta_1+ \Delta_2 -\Delta_{12}}{2} & \frac{\Delta_2 + \Delta_4 -\Delta_{12}-\Delta_{23}}{2}\\
        \frac{\Delta_1+ \Delta_2 -\Delta_{12}}{2} & \Delta_2 & \frac{\Delta_2+ \Delta_3 -\Delta_{23}}{2} \\
        \frac{\Delta_2 + \Delta_4 -\Delta_{12}-\Delta_{23}}{2} & \frac{\Delta_2+ \Delta_3 -\Delta_{23}}{2}  & \Delta_3
    \end{pmatrix}
\end{equation}
must be positive semi-definite.\footnote{This in particular implies the following weaker inequalities
\begin{equation}
\begin{split}
      &(\sqrt{\Delta_1} -\sqrt{\Delta_2}  )^2   \leq \Delta_{12}\leq (\sqrt{\Delta_1} +\sqrt{\Delta_2}  )^2, \\
      &(\sqrt{\Delta_2} -\sqrt{\Delta_3}  )^2   \leq \Delta_{23}\leq (\sqrt{\Delta_2} +\sqrt{\Delta_3}  )^2
\end{split}
\end{equation}
}
One can indeed verify that all our previous examples give semi-positive-definite covariances.

We may also consider more general constructions such as the ones discussed in section \ref{sec:general-lattice},  which allow for more terms to contribute, potentially allowing more matches with Witten diagram results.  We will not do this exercise here.

\section{Discussion}
We have shown how to build contact diagrams from chords and provided a microscopic basis for such constructions---models with random fluxes in their Fock spaces. 
Even in the simplest models we  constructed,  we found diagrams that give matching results with some of the scalar AdS$_2$ contact Witten diagrams,  which is especially nontrivial for four-point functions. This further lends support to the thesis that chords may represent spacetime processes in (near-)AdS$_2$.  Our work potentially provides a starting point for building up the notion of locality from chords: the leading-order chord diagrams are sufficient to produce the Schwarzian/Jakiw-Teitelboim dynamics, but such dynamics can be entirely described as the boundary fluctuations of the AdS$_2$ geometry, and it is hard to imagine how the notion of bulk locality can arise from this.  We showed how certain subleading chord diagrams reproduce contact interactions in the bulk, and the contactness seems like a natural representation of locality.  

We may also ponder whether this provides any lesson on how to connect our plethora of near-CFT$_1$ models to the UV, such as $\mathcal{N}=4$ super Yang-Mills. As have been elaborated in previous works \cite{Berkooz_2024,berkooz2023parisis,Jia:2024tii} and briefly alluded to in the current paper, it is almost too easy to invent a model that produces the Schwarzian dynamics within the context of fluxed Fock space models:  as long as you can build some fluxed operators  with some mild constraints ($p$-fermions, $p$-spins, $T^\pm_\mu$ operators etc.), then you can combine them every which way to form Hamiltonians and probe operators --- very little requirement is needed on the index sets of these operators --- most of them will give you the right Schwarzian dynamics. In this work we saw that if more  constraints are imposed, such as the ability to reproduce bulk contact correlations, we will need to choose more specific forms of the index sets and we need to impose specific  correlations on the fluxes (reflected by the quantities $\Delta_{ij}$ in section \ref{sec:fourptfunc}).  From a UV perspective, perhaps such qualities are should be expected in the first place.  For example, one may fantasize that our Fock space oscillators really come from the $psu(2,2|4)$ oscillators of $\mathcal{N}=4$ super Yang-Mills,  and that the random fluxes come from Berry phase mechanism after we adiabatically reduce the strongly-coupled super Yang-Mills (in a highly excited, thermalized state) to the infrared. In other words, since the strongly-coupled super Yang-Mills is described by very particular spin-chain Hamiltonians \cite{Beisert_2011}, we should really look for fluxed-deformations of these particular spin-chain Hamiltonians. And whatever structures these spin-chain Hamiltonians carry, they must be inherited by the fluxed models, including but not limited to the index sets structure.   
\acknowledgments{
  We  thank Micha Berkooz, Suman Kundu and Ohad Mamroud for discussions. This work is funded by an Israel Science Foundation center for excellence grant (grant number 2289/18), by grant no. 2018068 from the United States-Israel Binational Science Foundation (BSF), by the Minerva foundation with funding from the Federal German Ministry for Education and Research, by the German Research Foundation through a German-Israeli Project Cooperation (DIP) grant ``Holography and the Swampland", by Koshland fellowship and by a research grant from Martin Eisenstein.
}\bibliographystyle{JHEP}
\bibliography{library}

\end{document}